\documentclass{aa} 
\usepackage[]{natbib}
\usepackage{graphicx}
\usepackage{txfonts}
\usepackage[colorlinks=true, citecolor=blue]{hyperref}
\newcommand{\orcid}[1]{\href{https://orcid.org/#1}{\includegraphics[width=10pt]{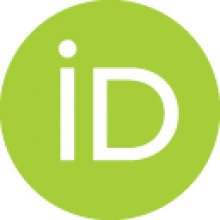}}}
\begin{document} 

\title{Galactic ArchaeoLogIcaL ExcavatiOns (GALILEO)} 
\subtitle{I. An updated census of APOGEE N-rich giants across the Milky Way}

\author{
Jos\'e G. Fern\'andez-Trincado\inst{1}\thanks{Correspondence should be addressed to:  jose.fernandez@ucn.cl and/or jfernandezt87@gmail.com}\orcid{0000-0003-3526-5052},
Timothy C. Beers\inst{2}\orcid{0000-0003-4573-6233}, 
Beatriz Barbuy\inst{3}\orcid{0000-0001-9264-4417},
Dante Minniti\inst{4,5}\orcid{0000-0002-7064-099X},
Cristina Chiappini\inst{6,7}\orcid{0000-0003-1269-7282}, 
Elisa R. Garro\inst{4}\orcid{0000-0002-4014-1591},
Baitian Tang\inst{8}\orcid{0000-0002-0066-0346},
Alan Alves-Brito\inst{9}\orcid{0000-0001-5579-2138},
Sandro Villanova\inst{10}\orcid{0000-0001-6205-1493},
Doug Geisler\inst{10,11,12}, 
Richard R. Lane\inst{13}\orcid{0000-0003-1805-0316}
\and
Danilo G. Diaz\inst{1}\orcid{0000-0003-1698-4924 }
}
        
\authorrunning{Jos\'e G. Fern\'andez-Trincado et al.} 
        
\institute{
	Instituto de Astronom\'ia, Universidad Cat\'olica del Norte, Av. Angamos 0610, Antofagasta, Chile
	\and
	Department of Physics and JINA Center for the Evolution of the Elements, University of Notre Dame, Notre Dame, IN 46556, USA
	\and
	Universidade de S\~ao Paulo, IAG, Rua do Mat\~ao 1226, Cidade Universit\'aria, S\~ao Paulo 05508-900, Brazil
	\and
	Depto. de Cs. F\'isicas, Facultad de Ciencias Exactas, Universidad Andr\'es Bello, Av. Fern\'andez Concha 700, Las Condes, Santiago, Chile
	\and
	Vatican Observatory, V00120 Vatican City State, Italy
	\and
	Leibniz-Institut f\"ur Astrophysik Potsdam (AIP), An der Sternwarte 16, 14482 Potsdam, Germany
	\and
	Laborat\'orio Interinstitucional de e-Astronomia - LIneA, Rua Gal. Jos\'e Cristino 77, Rio de Janeiro, RJ - 20921-400, Brazil
	\and
	School of Physics and Astronomy, Sun Yat-sen University, Zhuhai 519082, China  
	\and
	Universidade Federal do Rio Grande do Sul, Instituto de F\'isica, Av. Bento Gon\c{c}alves 9500, Porto Alegre, RS, Brazil        
	\and
	Departamento de Astronom\'\i a, Casilla 160-C, Universidad de Concepci\'on, Concepci\'on, Chile 
	\and
	Department of Astronomy - Universidad de La Serena - Av. Juan Cisternas, 1200 North, La Serena, Chile
	\and
	Instituto de Investigaci\'on Multidisciplinario en Ciencia y Tecnolog\'ia, Universidad de La Serena. Benavente 980, La Serena, Chile    
	\and
	Centro de Investigaci\'on en Astronom\'ia, Universidad Bernardo O'Higgins, Avenida Viel 1497, Santiago, Chile
}
        
        \date{Received ...; Accepted ...}
        \titlerunning{An Updated Census of N-rich stars Across the Milky Way}
        
        
        \abstract
        {We use the 17th data release of the second phase of the Apache Point Observatory Galactic Evolution Experiment (APOGEE-2) to provide a homogenous census of N-rich red giant stars across the Milky Way (MW). We report a total of 149 newly identified N-rich field giants toward the bulge, metal-poor disk, and halo of our Galaxy. They exhibit significant enrichment in their nitrogen abundance ratios ([N/Fe] $\gtrsim+0.5$), along with simultaneous depletions in their [C/Fe] abundance ratios ([C/Fe] $< +0.15$), and they cover a wide range of metallicities ($-1.8 < $ [Fe/H] $ <-0.7$). The final sample of candidate N-rich red giant stars with globular-cluster-like (GC-like) abundance patterns from the APOGEE survey includes a grand total of $\sim$ 412 unique objects. These strongly N-enhanced stars are speculated to have been stripped from GCs based on their chemical similarities with these systems. Even though we have not found any strong evidence for binary companions or signatures of pulsating variability yet, we cannot rule out the possibility that some of these objects were members of binary systems in the past and/or are currently part of a variable system. In particular, the fact that we identify such stars among the field stars in our Galaxy provides strong evidence that the nucleosynthetic process(es) producing the anomalous [N/Fe] abundance ratios occurs over a wide range of metallicities. This may provide evidence either for or against the uniqueness of the progenitor stars to GCs and/or the existence of chemical anomalies associated with likely tidally shredded clusters in massive dwarf galaxies such as ``Kraken/Koala,'' \textit{Gaia}-Enceladus-Sausage, among others, before or during their accretion by the MW. A dynamical analysis reveals that the newly identified N-rich stars exhibit a wide range of dynamical characteristics throughout the MW, indicating that they were produced in a variety of Galactic environments.
        }

        \maketitle
        
\section{Introduction}

The formation and assembly history of the Milky Way (MW) reveal that during early epochs it was an inexhaustible galactic cannibal, which has completely devoured a number of other large and massive dwarf galaxies some 9 to 11 Gyr ago \citep[see e.g.,][]{Kruijssen2020, Helmi2020}. Some of these now defunct dwarf galaxies have been identified as stellar structures or debris that populate the present-day Galactic field (e.g., the bulge, disk, and/or halo of the MW), and they are very phase mixed. Among the most populated progenitors associated to these systems are the Helmi Stream \citep{Helmi1999,Chiba2000}, the \textit{Gaia}-Sausage \citep{Belokurov2018} also referred to as  \textit{Gaia}-Enceladus \citep{Helmi2018}, Sequoia \citep{Myeong2019, Barba2019}, Thamnos \citep{Koppelman2019}, Kraken/Koala \citep{Kruijssen2020, Forbes2020}, and Pontus \citep{Malhan2022}, which serve as the fossil records of the events that created them, acting as time capsules that preserve the dynamical and chemical memory of the MW assembly history. \newline

Besides the ancient accretion events, there is a wealth of observational evidence for a more recent \citep[$<$ 7 Gyr ago; see e.g.,][]{Kruijssen2020} event which corresponds to the Sagittarius dwarf spheroidal galaxy \citep{Ibata1994, Hasselquist2017, Hasselquist2019} and its dispersed streams \citep{Hasselquist2017, Hasselquist2019}, which also preserve traces of tidal stripping and disruption of globular clusters (GCs) \citep[see e.g.,][]{Fernandez-Trincado2021_M54}, followed by the unusual Omega Centauri (NGC~5139) system \citep{Villanova2014}, together with its hypothesized ``Fimbulthul" stream \citep{Ibata2019} as well as its controversial tidal tails and stellar debris \citep{Costa2008, Majewski2012, Trincado2015a, Trincado2015b,  Marconi2014, Sollima2020}, and a number of stellar streams with variate origins and over a wide range of metallicities \citep{Wan2020, Ibata2021, Martin2022}. These claims are bolstered by the detection of extratidal stars with unusual GC-like abundance patterns, similar to that seen toward the Large and Small Magellanic systems today \citep{Fernandez-Trincado2020_MCs}, the stellar halo \citep{Hanke2020}, and the Galactic bulge \citep{Minniti2018, Fernandez-Trincado2021_NGC6723}. 

In addition to the major accretion events, the Galactic field is also polluted by a plethora of minor substructures \citep{Koppelman2019}, some of which have been named Aleph, Arjuna, I’itoi, LMS-1 (Wukong), among other \citep[see e.g.,][]{Naidu2020, Malhan2022}.  These substructures are in turn accompanied by a number of chemically atypical stars associated with dissolved GCs, such as the genuine population of giants with GC second-generation-like chemical patterns \citep{Fernandez-Trincado2016_2G_GCs, Fernandez-Trincado2017_2G} identified toward the bulge, disk, and inner halo \citep{Fernandez-Trincado2020_Orbits}. The recently discovered ``\texttt{Jurassic}” substructure \citep{Fernandez-Trincado2019_NewSirich, Fernandez-Trincado2020_Jurassic} is accompanied by a prominent population of aluminum-enhanced stars buried in the inner Galaxy \citep{Fernandez-Trincado2020_Alrich, Lucey2022}.  There is also a recently discovered population of high-metallicity ([Fe/H]$>-0.7$) giants with atmospheres strongly enhanced in nitrogen and aluminum \citep{Fernandez-Trincado2021_Hmetallicity} well above typical Galactic levels. 

All of these chemically unusual populations were likely formed after the infall of major satellites and/or stellar clusters during different epochs of MW evolution, as envisioned by many studies \citep[see e.g.,][and references therein]{Leon2000, Majewski2012, Kunder2014, Anguiano2015, Trincado2015b, Trincado2015a, Fernandez-Trincado2016_47Tuc, Fernandez-Trincado2016_2G_GCs, Lind2015, Recio-Blanco2017, Minniti2018, Kundu2019a, Kundu2019b, Hanke2020, Sollima2020, Kundu2021}.  
Based on low- and high-resolution spectroscopic surveys, several pioneering works have searched for traces of a unique and exclusive collection of field stars that contrast with the typical chemical composition of the majority of Galactic stars, and they are uniquely distinctive in some of the canonical abundance planes and speculated to belong to the population of GCs' escapee. Among these field stars are the nitrogen-enhanced metal-poor stars \citep[NEMP;][]{Johnson2007} with [C/N] $<-0.5$ and [N/Fe] $> +0.5$, which have also been exclusively identified in GC environments \citep[see e.g.,] []{Simpson2019}.

Previous studies along these lines include a unique collection of carbon-deficient ([C/Fe] $<+0.15$) nitrogen-enhanced field stars (hereafter N-rich stars) identified toward the inner Galaxy by \citet{Fernandez-Trincado2016_2G_GCs} and \citet{Martell2016}; toward the bulge field by \citet{Schiavon2017}, \citet{Fernandez-Trincado2017_2G}, and \citet{Fernandez-Trincado2019_Chemodynamics}; and an extra-tidal N-rich star from the bulge GC NGC~6723 was identified by \citet{Fernandez-Trincado2021_NGC6723}. A number of more recently added examples are from \citet{Kisku2021} in the inner Galaxy, toward the disk field, and beyond the tidal radius of Palomar~5 by \cite{Fernandez-Trincado2019_Chemodynamics}, and in the field of the Large and Small Magellanic Clouds (LMC and SMC, respectively) and the Sagittarius Spheroidal galaxy by \citet{Fernandez-Trincado2020_MCs} and \citet{Fernandez-Trincado2021_M54}, respectively. Additionally, the similarity in orbital properties between GCs and N-rich field stars (or previously called ``CN-strong field stars") was demonstrated in \citet{Carollo2013} and \citet{Tang2020}. While the wide gamut of elemental abundances recently examined by \citet{Yu2021} with optical spectra for some of these N-rich stars strengthened their exclusive link to GCs.

\begin{figure*}
	\begin{center}
		\includegraphics[width=190mm]{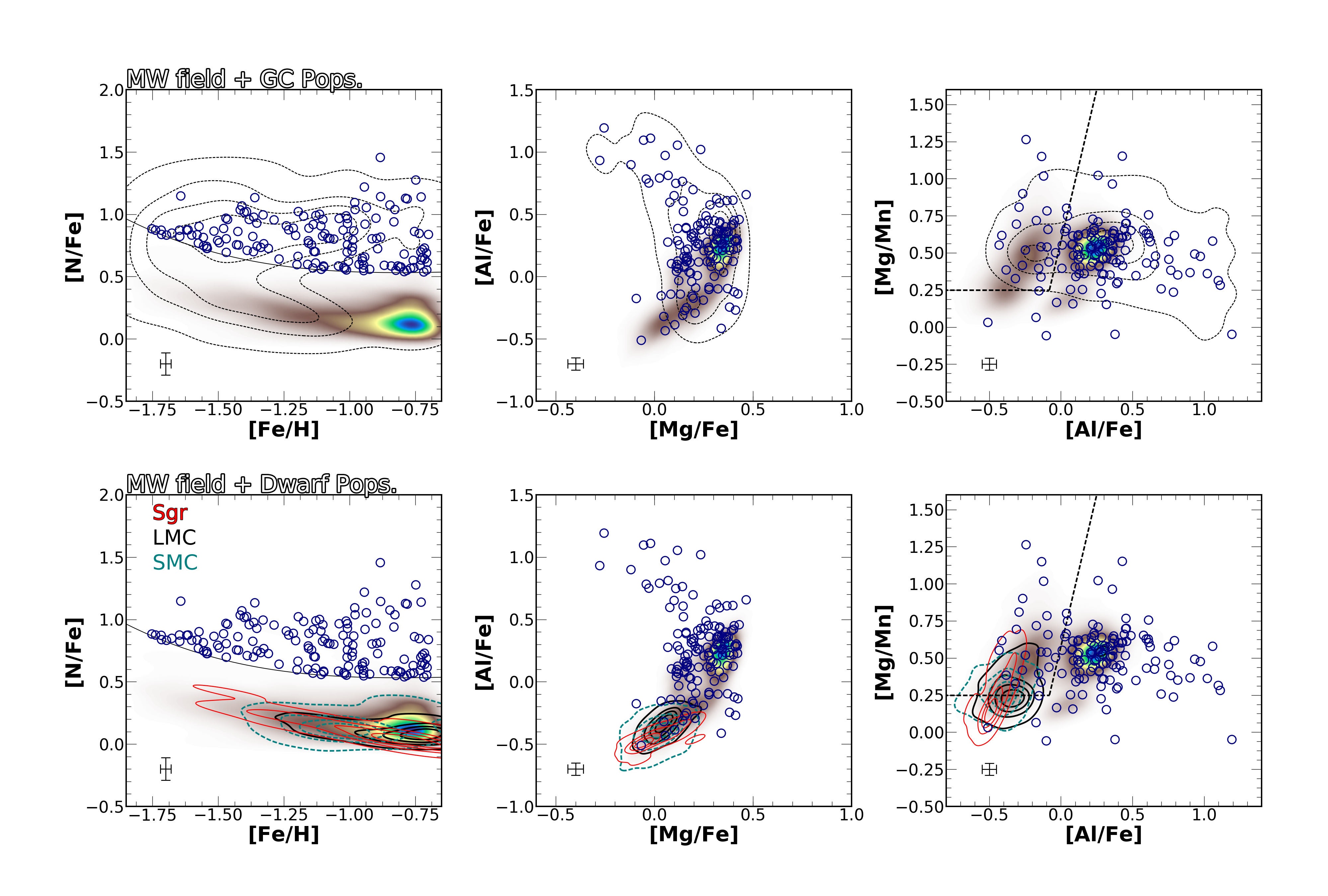}
		\caption{Distribution of [N/Fe] vs. [Fe/H], [Al/Fe] vs. [Mg/Fe], and [Mg/Mn] vs. [Al/Fe] for the newly identified N-rich stars (navy) compared to GC stars (black dashed contours) from \citet{Meszaros2020}, LMC (solid black contours), SMC (cyan dashed contours), and Sgr (solid red contours) stars from \citet{Helmi2018} with \texttt{ASPCAP} abundance information. The black dashed lines roughly define the criterion to ``distinguish"  in situ from accreted populations, similar to that as defined in \citet{Das2020} and \citet{Horta2021}. Also included are kernel density estimation (KDE) models showing the density of objects belonging to the MW. The continuous black line in the first column represents the fourth order polynomial fit at the 4$\sigma$ level selection criterion for N-rich stars as described in the text. Typical error bars are shown as black plus symbols in each panel.}
		\label{Figure1}
	\end{center}
\end{figure*}

Besides the carbon-deficient N-rich field stars, there is a significant population of  carbon-deficient aluminum-enriched  field stars (hereafter Al-rich stars) that exhibit low and high levels of nitrogen abundance ratios \citep{Fernandez-Trincado2020_Alrich}, suggesting that there are at least two groups of N-rich field stars that have likely been formed from two different nucleosynthetic pathways over a wide range of metallicities. Such anomalous abundance patterns are found in many GCs over a wide range of metallicities \citep{Fernandez-Trincado2019NGC6522, Fernandez-Trincado2020UKS1, Fernandez-Trincado2021Ton1, Meszaros2020, Meszaros2021, Geisler2021, Romero2021, Fernandez-Trincado2021patchick125, Fernandez-Trincado2021Ton2, Masseron2019}, and even into the most metal-poor MW GC \citep[see e.g.,][]{Fernandez-Trincado2021VVVCL}. Thus, the escape of stars from GCs could be a plausible explanation for the presence of these chemically anomalous stars in the Galactic field.

Although the recognized number of such chemically anomalous field stars has expanded greatly during the last decades, the origin of their unusual abundance signatures remains under discussion. For instance, recently, \citet[][]{Bekki2019} proposed an alternative way to produce these N-rich stars, which could be formed in high-density environments. More recently, \citet{Fernandez-Trincado2019_binary} has confirmed the existence of a N-rich star in a genuine binary system, which owes its nitrogen enrichment to binary mass transfer of an evolved asymptotic giant branch (AGB) star that is likely now a white dwarf. There still remains the possibility that a high fraction of the N-rich and Al-rich field stars could be evolved objects, possibly an "early-AGB" or an AGB star, rather than being formed through GC self-enrichment \citep[see e.g.,][]{Pereira2017}. The presence of such a young and moderately metal-poor star in the Galactic field would have important implications. However, its nature remains uncertain.

As emphasized by \citet{Tang2019}, some of the CH-normal field stars and the N-rich field stars could have a very similar origin, and likely they share the same nucleosynthetic pathways. The wide range of elemental abundances recently examined by \citet{Jincheng2021} for some of these N-rich stars has strengthened the proposed link to GCs. Thus, future dedicated long-term radial-velocity monitoring of all of these stars would naturally be the best course of action in order to offer clues about the rare astrophysical events behind these unique and atypical objects in the MW. 

The zoo of chemically anomalous stars in the Galactic field is very diverse. In addition to the previously mentioned N-rich and Al-rich stars, substantial evidence has been offered for the existence of Si-rich stars \citep{Fernandez-Trincado2019_NewSirich, Fernandez-Trincado2020_Jurassic}, K-rich stars \citep{Kemp2018}, \textit{r}-process enriched stars \citep{Holmbeck2020, Gudin2021, Shank2021}, \textit{s}-process enriched stars \citep{Anders2018, Pereira2019Srich}, Na-rich stars \citep{Pereira2019Na, Koo2021}, and P-rich stars \citep{Masseron2020a, Masseron2020b} over a wide range of metallicities, and they are thought to form through different nucleosynthetic channels.

Here, we report the results of the analysis of the final and complete data release of the Apache Point Observatory Galactic Evolution Experiment II survey \citep[APOGEE-2;][]{Majewski2017} through Galactic ArchaeoLogIcaL ExcavatiOns (GALILEO). This project dedicated to searching for and detecting chemically anomalous giants -- including the stars that show anomalously high levels of [N/Fe] and [Al/Fe] -- for long-term radial-velocity monitoring to better understand the nature of these stars with unusual elemental abundances, as well as its nucleosynthetic pathways.

\section{Observations and data}

We employed high-resolution ($R\sim22,500$) near-infrared (NIR) spectra taken from the final APOGEE-2 seventeenth data release \citep[DR~17;][]{DR17}, which is one of the internal programs of the Sloan Digital Sky Survey-IV \citep{Blanton2017}.\ These programs have been developed to provide precise radial velocities (RVs)  $<$1 km s$^{-1}$ and elemental abundances for over 700,000 unique stars across the MW and nearby systems.

APOGEE-2 observations were carried out through two identical spectrographs \citep{Wilson2019} from the Northern Hemisphere on the 2.5m telescope at Apache Point Observatory \citep[APO, APOGEE-2N;][]{Gunn2006} and from the Southern Hemisphere on the Ir\'en\'ee du Pont 2.5m telescope \citep[][]{Bowen1973} at Las Campanas Observatory (LCO, APOGEE-2S). Each instrument records most of the \textit{H} band (1.51$\mu$m -- 1.69$\mu$m) on three detectors, with coverage gaps between $\sim$1.58--1.59$\mu$m and $\sim$1.64--1.65$\mu$m, and with each fiber subtending a $\sim$2'' diameter on-sky field of view for the northern instrument and 1.3'' for the southern one. The targeting strategy and plan for the updates of the APOGEE-2 survey is fully described in \citet{Zasowski2013}, \citet{Zasowski2017}, and \citet{Beaton2021} for Northern Hemisphere observations, and in \citet{Santana2021} for Southern Hemisphere observations.

The APOGEE-2 spectra have been reduced as described in \citet{Nidever2015}, \citet{Holtzman2018}, and \citet{Jonsson2018}, and analyzed using the APOGEE Stellar Parameters and Chemical Abundance Pipeline \citep[\texttt{ASPCAP};][]{Garcia2016}, as well as the libraries of synthetic spectra described in \citet{Zamora2015}. The customized \textit{H}-band line lists are fully described in \citet{Shetrone2015}, \citet{Hasselquist2016} (neodymium lines,  Nd II), \citet{Cunha2017} (cerium lines, Ce II), \citet{Masseron2020} (phosphorus lines), and have been recently updated and reviewed in \citet{Smith2021}. 

\section{Data selection}
\label{quality}

Our goal was to search for the ambiguous population of N-rich stars that likely detached from GCs across the MW by making use of the latest data release of the APOGEE-2 survey. Before selecting potential N-rich stars, we were required to apply a number of criteria to certify the quality of the data for our analysis. We first cleaned the sample from sources with unreliable parameters by removing all stars with \texttt{STARFLAG} and \texttt{ASPCAPFLAG} different from zero \citep[see e.g.,][]{Holtzman2015}. That is to say, \texttt{ASPCAPFLAG} $==0$ ensures that there are not major flagged issues, such as a low signal-to-noise ratio (S/N), poor synthetic spectral fit, stellar parameters near grid boundaries, among others; whilst \texttt{STARFLAG} $==0$ minimizes potentially problematic object spectra. We then limited the data to red giant stars with spectral S/Ns larger than 70 pixel$^{-1}$, stellar effective temperatures in the range 3200\,K -- 5500\,K, surface gravities $\log g <$ 3.6, and metallicities ([Fe/H]) between $-1.8$ and $-0.7$, so as to minimize the presence of sources from the thin and thick disk at the metal-rich end, and to include giants with reliable carbon and nitrogen abundances at the metal-poor end. We also removed sources with C\_FE\_FLAG $==$ {BAD}, N\_FE\_FLAG $==$ {BAD}, O\_FE\_FLAG $==$ {BAD}, and FE\_H\_FLAG $==$ {BAD}.

Following the conventions in the literature \citep[see e.g.,][]{Fernandez-Trincado2016_2G_GCs, Fernandez-Trincado2019_Chemodynamics}, we required noncarbon-enhanced stars with [C/Fe] $< +$0.15, which are typically found in GC environments. These criteria allowed us to reduce the presence of possible CH stars \citep{Karinkuzhi2015}. 

Next, we removed a total of 1826 stars from the sample which are deemed to be potential members of GCs, as listed in the the catalogs of \citet{Masseron2019}, \citet{Meszaros2020}, and \citet{Meszaros2021}, and in the more recent compilation from Baumgardt's web service\footnote{\url{https://people.smp.uq.edu.au/HolgerBaumgardt/globular/}} which is fully described in \citet{Vasiliev2021} and \citet{Baumgardt2021}. We also removed N-rich red giant stars previously reported in the literature \citep[see e.g.,][]{ Fernandez-Trincado2016_2G_GCs, Fernandez-Trincado2017_2G, Fernandez-Trincado2019_Chemodynamics, Fernandez-Trincado2019_NewSirich, Fernandez-Trincado2019_binary, Fernandez-Trincado2020_Alrich, Fernandez-Trincado2020_MCs, Fernandez-Trincado2020_Jurassic, Fernandez-Trincado2021_NGC6723, Fernandez-Trincado2021_M54, Martell2016, Schiavon2017, Kisku2021}.

The application of the above filters and removing duplicate entries left us with a sample of 8812 unique sources with high-quality spectral information. The data for the objects of interest were extracted from this final sample as described in the following sections. 

\section{Newly identified N-rich stars}
\label{new}

We started with a nitrogen- and metallicity-based selection criterion in the [Fe/H] versus [N/Fe] plane, which is presented in Figure \ref{Figure1} (\textit{top-left} panel), and similar to the one used in \citet{Martell2016} and \citet{Fernandez-Trincado2019_Chemodynamics}. We fit a fourth-order polynomial to the distribution in the [N/Fe]--[Fe/H] plane which captures the mean behavior of our data set, and we selected all stars more than 4$\sigma$ above that curve as N-rich. As can be seen in the \textit{top-left} panel in  Figure \ref{Figure1}, the newly identified N-rich stars exhibit typical nitrogen abundance ratios of [N/Fe] $> +0.5$. This returns a large sample of 149 newly identified unique red giant stars that are strongly nitrogen-enhanced relative to the final data set. The main kinematic properties, orbital elements, and elemental abundances of these newly identified N-rich stars are listed in Table \ref{Table1}.

Table \ref{Table1} also lists five newly identified N-rich star candidates toward the SMC (three stars: 2M01055438$-$6833068, 2M01021954$-$7147493, and 2M00592789$-$6943291) and LMC (two stars: 2M05023164$-$7157181 and 2M05481230$-$7937052), which display similar physical properties to that seen in the stars of these systems. We do not provide orbital calculations for these stars because of the reasons provided in Section \ref{dynamics}. These stars have been cataloged as \texttt{Orbital\_Sense} $==$ ``\texttt{Unknow/LMC-Candidate}'' and \texttt{Orbital\_Sense} $==$ ``\texttt{Unknow/SMC-Candidate}''. 

N-rich stars in Table \ref{Table1} with \texttt{Orbital\_Sense}$==$``\texttt{prograde/HVNS-Candidate,}'' have been classified as hypervelocity N-rich star (HVNS) candidates. While N-rich stars in our sample with unreliable orbital solutions have been tagged as \texttt{Orbital\_Sense}$==$``\texttt{Unknow},'' stars in different Galactic orbital configurations have been classified as \texttt{Orbital\_Sense}$==$``\texttt{P--R}'' to refer to the orbits changing their sense of motion from prograde to retrograde (P--R) during the integration time and vice versa, \texttt{as Orbital\_Sense}$==$``\texttt{Prograde}'' to highlight those stars in a prograde orbital configuration, and as \texttt{Orbital\_Sense}$==$``\texttt{Retrograde}'' to refer to those that lie in retrograde orbits. We characterize this newly identified N-rich population in Section \ref{dynamics}.

\begin{table*}
        \begin{small}
                \begin{center}
                        \setlength{\tabcolsep}{2.5mm}  
                        \caption{Description of the columns of the catalog containing the 149 newly identified N-rich stars. This table is published in its entirety as supporting information with the electronic version of the article.
                        }
                        \begin{tabular}{llll}
\hline
\hline          
ID    & Column name  & Units       & Column Description \\
\hline
\hline
1       &       APOGEE\_ID     &      &   APOGEE id \\
2       &       RA                        & degrees   &   $\alpha$ (J2000)     \\
3       &       DEC                       & degrees   &     $\delta$ (J2000)        \\
4       &       J\_2MASS           &    &        2MASS J band    \\
5       &       H\_2MASS           &    &        2MASS H band    \\
6       &       K\_2MASS           &    &        2MASS K band    \\
7       &       BP\_GaiaEDR3           &    &        \textit{Gaia} EDR3 BP band    \\
8       &       RP\_GaiaEDR3           &    &        \textit{Gaia} EDR3 RP band    \\
9       &       G\_GaiaEDR3           &    &        \textit{Gaia} EDR3 G band    \\
10      &        SNR           &           pixel$^{-1}$    & spectral signal-to-noise \\
11      &       RV           &           km s$^{-1}$    &  APOGEE-2 radial velocity \\
12      &       VSCATTER           &           km s$^{-1}$    &  APOGEE-2 radial velocity scatter \\
13      &       TEFF            &      K  & effective temperature \\
14      &       ERROR\_TEFF  &      K  & uncertainty in TEFF\\
15      &       LOGG &      [cgs]  & surface gravity \\
16      &       ERROR\_LOGG  & [cgs]  & uncertainty in LOGG \\
17      &       C\_FE  &  &  [C/Fe] from \texttt{ASPCAP} \\
18      &       ERROR\_C\_FE &  & uncertainty in [C/Fe] \\
19      &       N\_FE  &  &  [N/Fe] from \texttt{ASPCAP} \\
20      &       ERROR\_N\_FE &  & uncertainty in [N/Fe] \\
21      &       MG\_FE  &  &  [Mg/Fe] from \texttt{ASPCAP} \\
22      &       ERROR\_MG\_FE &  & uncertainty in [Mg/Fe] \\
23      &       AL\_FE  &  &  [Al/Fe] from \texttt{ASPCAP} \\
24      &       ERROR\_AL\_FE &  & uncertainty in [Al/Fe] \\
25      &       MN\_FE  &  &  [Mn/Fe] from \texttt{ASPCAP} \\
26      &       ERROR\_MN\_FE &  & uncertainty in [Mn/Fe] \\
27      &       FE\_H  &  &  [Fe/H] from \texttt{ASPCAP} \\
28      &       ERROR\_FE\_H &  & uncertainty in [Fe/H] \\
29      &  VR       & km s$^{-1}$ & Galactocentric radial \\
        &        & &  velocity in V$_{\rm R}$ \\
30      & ERROR\_VR  & km s$^{-1}$ & uncertainty in V$_{\rm R}$  \\
31      & VPHI   & km s$^{-1}$ & Galactocentric azimuthal \\
        &    & &  velocity in V$_{\rm \phi}$ \\
32       & ERROR\_VPHI  & km s$^{-1}$ & uncertainty in V$_{\rm \phi}$  \\
33      & PERIGALACTICON &  kpc & Perigalactocentric distance\\
34      & ERROR\_PERIGALACTICON & kpc & uncertainty in PERIGALACTICON\\
35      & APOGALACTICON & kpc & Apogalactocentric distance\\
36      & ERROR\_APOGALACTICON & kpc &  uncertainty in APOGALACTICON\\
37      & Eccentricity   & & Orbital eccentricity\\
38      & ERROR\_Eccentricity & & uncertainty in Eccentricity\\
39      & Zmax &  kpc &  Maximum vertical excursion\\
        &  &   &   from the Galactic plane\\
40      & ERROR\_Zmax & kpc & uncertainty in Zmax\\
41      & Orbital\_Sense  &  & Orbital sense\\
42      & ruwe & &  \textit{Gaia} EDR3 renormalized \\
        &  & &  unit weight error \\
43      & d\_STARHORSE$^{\dagger}$ &kpc & Bayesian \texttt{StarHorse} distance, 50th percentile \\
44      & ERROR\_d\_STARHORSE & kpc  & uncertainty in d\_STARHORSE , \\
        &  &   & ERROR $=$ (84th - 16th)/2 percentile \\
45      & pmRA\_GaiaEDR3  & mas yr$^{-1}$ &  $\mu_{\alpha}\cos{}(\delta)$ from \textit{Gaia} EDR3\\
46      & ERROR\_pmRA\_GaiaEDR3 & mas yr$^{-1}$ & uncertainty in  $\mu_{\alpha}\cos{}(\delta)$  \\
        & &  & from \textit{Gaia} EDR3 \\
47      & pmDEC\_GaiaEDR3 & mas yr$^{-1}$  & $\mu_{\delta}$ from \textit{Gaia} EDR3\\
48      & ERROR\_pmDEC\_GaiaEDR3 & mas yr$^{-1}$  &  uncertainty in $\mu_{\delta}$ from \textit{Gaia} EDR3\\
                                \hline
                                \hline
                        \end{tabular}  \label{Table1}
                \end{center}
        \raggedright{{\bf Note:}  $^{\dagger}$For the stars 2M17025992$-$3537464 and 2M19175783$-$1343049, we provide the estimated Bayesian \texttt{StarHorse} and \texttt{Bailer-Jones} distances and their resulting parameters  in two different entries of the catalog.}      
        \end{small}
\end{table*}   

It is also important to note that none of the 149 N-rich candidates without strong carbon enhancement have a particularly strong variability in their radial velocity over the period of the APOGEE-2 observations, thus limiting the possible orbital properties than can be detected in our data.  Those sources with available multi-epoch observations exhibit a typical V$_{\rm scatter} < 0.9 $ km s$^{-1}$, with the exception of three sources, which show a slightly high V$_{\rm scatter}$ between 1.64 -- 2.6 km s$^{-1}$. We also found no matches with the ASAS-SN Catalog of Variable Stars \citep{Jayasinghe2021}. 

Although we have not found strong evidence for a binary companion or a pulsating variable star as of yet, we cannot rule out the possibility that some of these stars were members of a binary system in the past \citep[see e.g.,][]{Fernandez-Trincado2019_binary}, or they could likely be part of the semi-regular (SR) variable populations \citep[see e.g.,][]{Fernandez-Trincado2020_MCs}. Therefore, long-term radial-velocity monitoring for all the identified N-rich stars is currently being carried out with the SDSS-V panoptic spectroscopic survey \citep{Kollmeier2017} which, with a baseline that is better ($\lesssim$ 6 months) than that of the APOGEE-2 DR~17 survey, would naturally be the best course to establish the number of such objects that formed through the binary channel or that currently belong to a kind of pulsating variable.

Table \ref{Table2} summarizes the nearly complete census in APOGEE of  N-rich field stars, for a grand total of $\sim$412 unique carbon-depleted N-rich candidates across the MW. It is expected that this number will increase further once the results from the new  Milky Way Mapper (MWM) program of SDSS-V \citep{Kollmeier2017} are published. Table \ref{Table2} does not include the recent exploration by \citet{Phillips2022}, which reports on the detection of two N-rich stars along the Palomar~5 Stream. These two stars were already identified previously by \citet{Martell2016} for 2M15204588$+$0055032, suggested as potential extra-tidal material of  Palomar~5 in \citet{Fernandez-Trincado2019_Chemodynamics} for 2M15183589$+$0027100, and confirmed as members of the Palomar~5 Stream by the \texttt{STREAMFINDER} algorithm in \citet{Ibata2021}. Table \ref{Table3} lists the APOGEE ids and origin on the literature of the 412 compiled APOGEE N-rich stars. 

Finally, we found that there are no known GCs within an angular separation of $1.5\times{} {\rm r_{tidal, GC}}$ arcmin out of the 149 newly reported N-rich stars in our sample, and there is not an association with the newly discovered stellar streams in the \textit{Gaia} universe \citet{Ibata2021}. However, some of them could be made up of dissipated GCs, which is investigated in detail by assessing other chemical species not accessible from the APOGEE-2 survey.   

\begin{table*}
        \begin{small}
                \begin{center}
                        \setlength{\tabcolsep}{1.0mm}  
                        \caption{Census updated of $\sim$412 unique field carbon-depleted N-rich ([C/Fe]$<+0.15$; [N/Fe]$\gtrsim+0.5$) stars  in the APOGEE survey}
                        \begin{tabular}{|l|l|c|c|l|}
\hline
   Number of  & Galactic component & Data release (DR) & Code &  Literature \\                              
      N-rich stars &  & &  &   \\                               
\hline
\hline
       5 & Halo: Migrants & DR~12 & \texttt{ASPCAP}  &   \citet{Martell2016} \\                              
\hline                                  
       1 & Halo &  DR~12 & \texttt{iSpec}$^{\rm (1)}$ and \texttt{MOOG}$^{\rm (2)}$ & \citet{Fernandez-Trincado2016_2G_GCs}\\
\hline
       58 & Bulge & DR~12 &  \texttt{ASPCAP}  & \citet{Schiavon2017}\\
 \hline      
       11 & Bulge, Disk and Halo & DR~13 &  \texttt{Turbospectrum}$^{\rm (3)}$  &\citet{Fernandez-Trincado2017_2G}\\
\hline       
        31 & Bulge, Disk, Halo, and& DR~14 & \texttt{Payne}  &\citet{Fernandez-Trincado2019_Chemodynamics}\\
                & Palomar~5: extra-tidal candidate& &  &\\
\hline       
         1 & Halo: Binary system  & DR~14 & \texttt{BACCHUS} &\citet{Fernandez-Trincado2019_binary}\\
\hline 
7 & Bulge, Disk and Halo: N-/Si-rich & DR~16 &\texttt{BACCHUS} &\citet{Fernandez-Trincado2020_Jurassic}\\               
& (with simultaneously C,N,O determinations) &&&\\                            
\hline
44 & Magellanic Clouds: LMC \& SMC  & DR~16 & \texttt{BACCHUS} &\citet{Fernandez-Trincado2020_MCs}\\           
\hline 
          16 & Bulge and inner Halo: N-/Al-rich & DR~16 &\texttt{BACCHUS} &\citet{Fernandez-Trincado2020_Alrich}\\               
\hline         
        47 out of 83& Located in the inner 4 kpc of the MW; & DR~16 & \texttt{ASPCAP}  &\citet{Kisku2021}\\
                 & excluding previously identified N-rich stars, and  & & &\\
                  & stars positioned inside the tidal radius of known GCs & & &\\
\hline
2 & Sgr$+$M~54 field & DR~16 & \texttt{BACCHUS} &\citet{Fernandez-Trincado2021_M54}\\   
\hline 
        40  out of 42 & Bulge, Disk and Halo ([Fe/H]$>-0.7$) & DR~17 & \texttt{ASPCAP}  &\citet{Fernandez-Trincado2021_Hmetallicity}\\    
         & excluding two previously identified N-rich stars & &&\\                        
\hline
        149 & LMC, SMC, Bulge, Disk and Halo ([Fe/H]$<-0.7$) & DR~17 & \texttt{ASPCAP} &{\bf This work}\\        
\hline
                        \end{tabular}  \label{Table2}
                \end{center}
\raggedright{{\bf Note:} $^{\rm (1)}$\citet{Blanco-Cuaresma2014}; $^{\rm (2)}$\citet{Sneden1973}; and $^{\rm (3)}$\citet{Alvarez1998}. }     
        \end{small}
\end{table*}   

\begin{table}
        \begin{small}
                \begin{center}
                        \setlength{\tabcolsep}{0.5mm}  
                        \caption{Compilation of the 412 N-rich stars indentified thus far in the APOGEE survey. The APOGEE id and literature source (doi) is listed. The newly identified N-rich stars are listed in this table as ``NEW." (This table is published in its entirety as supporting information with the electronic version of the article. A portion is shown here for guidance regarding its form and content).}
                        \begin{tabular}{cc}
\hline
\hline
APOGEE\_ID          &   doi                         \\
\hline 
\hline
  2M16011638$-$1201525 &   10.3847/1538-4357/833/2/132 \\
  2M16051144$-$2330557 &   10.1093/mnras/stab525       \\
  2M16100035$-$1933027 &   10.1093/mnras/stab525       \\
  2M16161605$-$2838569 &   10.1093/mnras/stab525       \\
  2M16180906$-$2442217 &   10.1093/mnras/stab525       \\
  2M16304650$-$2949522 &   10.1093/mnras/stab525       \\
  2M16314726$-$2945273 &   10.1093/mnras/stab525       \\
  2M16333703$-$3028333 &   10.1093/mnras/stab525       \\
  2M16335569$-$1344044 &   10.1093/mnras/stab525       \\
  2M16493016$-$2557027 &   10.1093/mnras/stab525       \\
  2M17043840$-$2608052 &   10.1093/mnras/stab525       \\
  2M17132833$-$1820552 &   10.1093/mnras/stab525       \\
  2M17281699$-$3024573 &   10.1093/mnras/stab525       \\
  2M17285196$-$2013080 &   10.1093/mnras/stab525       \\
  2M17292082$-$2126433 &   10.1093/mnras/stab525       \\
  2M17293012$-$3006008 &   10.1093/mnras/stab525       \\
  2M17293730$-$2725594 &   10.1093/mnras/stab525       \\
  2M17305645$-$3030155 &   10.1093/mnras/stab525       \\                               
  ... & ... \\
  2M19175783$-$1343049   & NEW \\
    ... & ... \\
                                \hline
                                \hline
                        \end{tabular}  \label{Table3}
                \end{center}
        \end{small}
\end{table}   

\section{Elemental abundance analysis}

In this section we analyze the chemical properties of the newly identified N-rich stars alongside the MW halo and disk, the LMC and SMC members, the Sagittarius spheroidal dwarf galaxy (Sgr), and GC members, in the chemical diagnostic plans of [N/Fe] versus [Fe/H], [Al/Fe] versus [Mg/Fe], and [Mg/Mn] versus [Al/Fe]. Our sample selection of LMC, SMC, and Sgr members was obtained by cross-matching the catalogs from Helmi's web service\footnote{\url{https://www.astro.rug.nl/~ahelmi/research/dr2-dggc/}} \citep{GaiaHelmi2018} against the APOGEE-2 DR~17 catalog \citep{DR17}, and adopting the quality cuts described in Section \ref{quality}. GC sources were adopted from \citet{Meszaros2020} and \citet{Meszaros2021}, but we adopted the APOGEE-2/\texttt{ASPCAP} DR~17 abundance determinations in order to avoid the systematic elemental abundances produced between \texttt{BACCHUS} \citep{Masseron2016} and \texttt{ASPCAP} \citep{Garcia2016}, as our sample relies on abundance ratios determined from the \texttt{ASPCAP} pipeline.

Figure \ref{Figure1} summarizes the elemental properties of the 149 newly identified N-rich red giant stars. The first column of the panels in this figure shows that the [N/Fe] abundance ratio of the N-rich stars is more compatible with the nitrogen-enriched GC population, rather than the LMC, SMC, and Sgr populations. 

The second column of panels in Figure \ref{Figure1} reveals that there appears to be at least two main groups in the [Mg/Fe] versus [Al/Fe] plane, where [Al/Fe] is anticorrelated with [Mg/Fe], similar to what is seen in GC populations, primarily for [Al/Fe] $\gtrsim$ 0, and where it also overlaps with the main bulk of ME stars. While the second group of N-rich stars with subsolar [Al/Fe]$<0$ fall within the group dominated by the dwarf members plus GC populations with low-aluminum enrichment more likely the first-generation of GC stars as envisioned by \citet[][]{Meszaros2020}. The third column of panels in Figure \ref{Figure1} shows  [Mg/Mn] versus [Al/Fe], as originally envisioned by \citet{Das2020}. The newly identified N-rich populations expand the same locus of that chemical plane deemed to contain GC populations, even (in)outside the ``accreted" population area defined by \citet{Das2020} which runs roughly above [Mg/Mn] $>+0.25$ and [Al/Fe] $\lesssim+0.2$. 

While \citet{Das2020} and \citet{Horta2021} suggest that the [Al/Fe]--[Mg/Mn] plane could be used to identify stellar populations that formed {ex situ}, some caution is recommended since metal-poor ([Fe/H] $\lesssim-0.7$) nitrogen-enriched populations that likely formed {ex situ} share the same locus  as the {in situ} MW population in this plane. In particular, for subsolar [Al/Fe], there is a mix of stellar populations belonging to GCs and dwarf populations. 

\section{Dynamical properties of selected N-rich stars}
\label{dynamics}

We examined the kinematic and dynamical properties of the newly identified N-rich stars by making use of the \texttt{GravPot16}\footnote{\url{https://gravpot.utinam.cnrs.fr}} model \citep{GravPot16}, and computing the ensemble of orbits integrated over a 3 Gyr time span. For the orbit calculations, we assumed a bar pattern speed of 41$\pm$10 km s$^{-1}$ kpc$^{-1}$ \citep{Sanders2019, Bovy2019}. We note that our model has some limitations in the processes considered; for example, secular changes in the adopted MW potential as well as dynamical friction are not included.

The orbit calculations were performed by adopting a simple Monte Carlo approach that considers the errors of the observables. The resulting values and their errors were taken as the 16$^{\rm th}$, 50$^{\rm th}$, and 84$^{\rm th}$ percentiles from the generated distributions. Heliocentric distances were estimated with the Bayesian \texttt{StarHorse} code \citep[see e.g.,][]{Anders2019, Queiroz2018, Queiroz2020, Queiroz2021} as part of the APOGEE\_STARHORSE Value Added Catalog\footnote{https://data.sdss.org/sas/dr17/env/APOGEE\_STARHORSE/}, proper motions are from \textit{Gaia} EDR~3 \citep{Brown2021}, and radial velocities are from the APOGEE-2 DR~17 database. 

Figure \ref{Figure2} shows the resulting kinematic properties and orbital elements for 124 out of 149 stars in our sample which have a \textit{Gaia} re-normalized unit weight error (\texttt{RUWE}) of less than 1.4 (indicating the high quality of their astrometric solutions), and uncertainties on the heliocentric distances of less than 20\%. It is important to note that five potential N-rich SMC and LMC members were excluded from our dynamical analysis; as our MW model does not take the Magellanic Clouds' contribution to the global gravitational potential into account, their orbital elements will be more uncertain.
 
\subsection{Kinematical properties} 

Figure \ref{Figure2} reveals that a significant fraction (65\%; 80 stars) lie within the fraction dominated by \textit{Gaia}-Enceladus-Sausage-like \citep[GES-like;][]{Belokurov2018, Helmi2018} stars and a halo-like population with $V_{\rm \phi}<100$ km s$^{-1}$, which exhibit a typical azimuthal velocity uncertainty below 40 km s$^{-1}$, with the exception of 2M16053258$-$4502140 exhibiting a relatively large azimuthal velocity uncertainty of $V_{\rm \phi} = -264 \pm  163$ km s$^{-1}$, but it is clearly well separated kinematically from the GES-like component and disk populations (see Figure \ref{Figure2}). While 34\% (42 stars) of the N-rich stars exhibit disk-like kinematics, with an azimuthal velocity of $V_{\rm \phi}>100$ km s$^{-1}$ with typical uncertainties $<35$ km s$^{-1}$ -- with the exception of 2M06243113$+$4213097 exhibiting a large uncertainty in $V_{\rm \phi} \sim 237\pm73$  km s$^{-1}$ -- they are still related to the disk-like kinematic population. However, as is described below, the newly identified N-rich stars cover a wide range of dynamical characteristics with overlapping Galactocentric radial and azimuthal velocities. 

\subsection{Dynamical properties} 

A more elaborate inspection of their orbital elements reveal that there are at least four main groups of N-rich stars in our sample, whose kinematic distributions are shown in Figure \ref{Figure3}. The first group (Group \#1) is dominated by a significant fraction (72 out of 124 stars) of N-rich stars which lie on pure prograde orbits with an azimuthal velocity of $V_{\rm \phi}\gtrsim-50$ km s$^{-1}$. 

A second group (Group \#2) is made up of 25 out of 124 N-rich stars which exhibit $V_{\rm \phi}<100$ km s$^{-1}$. They also have a retrograde orbital configuration with respect to the rotation of the Galactic bar. 

A third group (Group \#3) composed of 25 out of 124 N-rich stars with Galactic azimuthal velocities ranging from $-150$ km s$^{-1}$ $\lesssim V_{\rm \phi}\lesssim 150$ km s$^{-1}$ exhibit an unusual dynamical behavior, that is to say they are in the so-called P--R orbital configuration which changes their sense of motion from prograde to retrograde during the integration time and vice versa, which could also be related to chaotic behavior. Further, for these three groups, we find no strong correlation between their orbital elements and [Fe/H], which could be indicative of an early chaotic phase of the evolution of the MW, as well. 

\begin{figure*}
        \begin{center}
                \includegraphics[width=190mm]{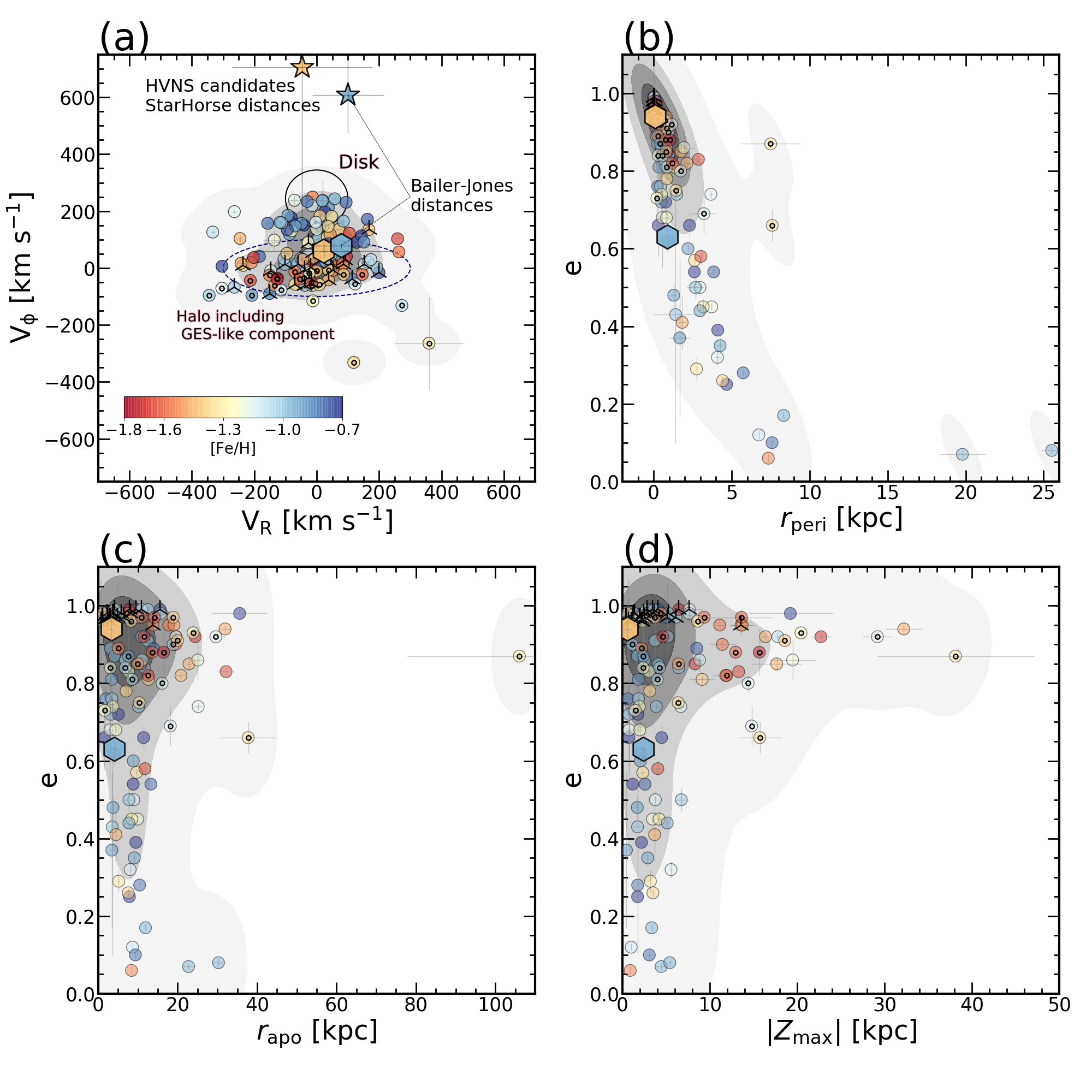}
                \caption{Galactocentric radial (V$_{\rm R}$) and azimuthal (V$_{\rm \phi}$) velocity, and orbital-element properties for the newly identified N-rich stars. The color-coding of the symbols in the panels is based on their [Fe/H], while the gray background shading represents the KDE model of the overlaid points, excluding the HVNS candidates (see text). The distribution of the velocity component V$_{\rm R}$ vs. V$_{\rm \phi}$ is shown in panel (a). The blue dashed line represents the approximate region for sources with the \textit{Gaia}-Enceladus-Sausage-like kinematics in V$_{\rm R}$ vs. V$_{\rm \phi}$, based on \citet{Belokurov2018}. The area occupied by disk-like stars is also highlighted with the empty black circle. The perigalactic (b), apogalactic distances (c), and the maximum vertical excursion (d) from the Galactic plane as a function of the orbital eccentricity are shown. Black up-tick symbols mark the stars with likely chaotic behavior, i.e., those with orbits that change their direction of motion from prograde to retrograde (and vice versa) with respect to the direction of Galactic rotation, while the small, empty black concentric symbols indicate the stars with retrograde orbits. Uncertainties are shown as gray plus symbols in each panel. The large star symbols mark the two HVNS candidates assuming the Bayesian \texttt{StarHorse} heliocentric distance estimations, while the large hexagonal markers refer to the same stars with \texttt{Bailer-Jones} heliocentric distance estimations.
                }
                \label{Figure2}
        \end{center}
\end{figure*}

\begin{figure}
        \begin{center}
                \includegraphics[width=90mm]{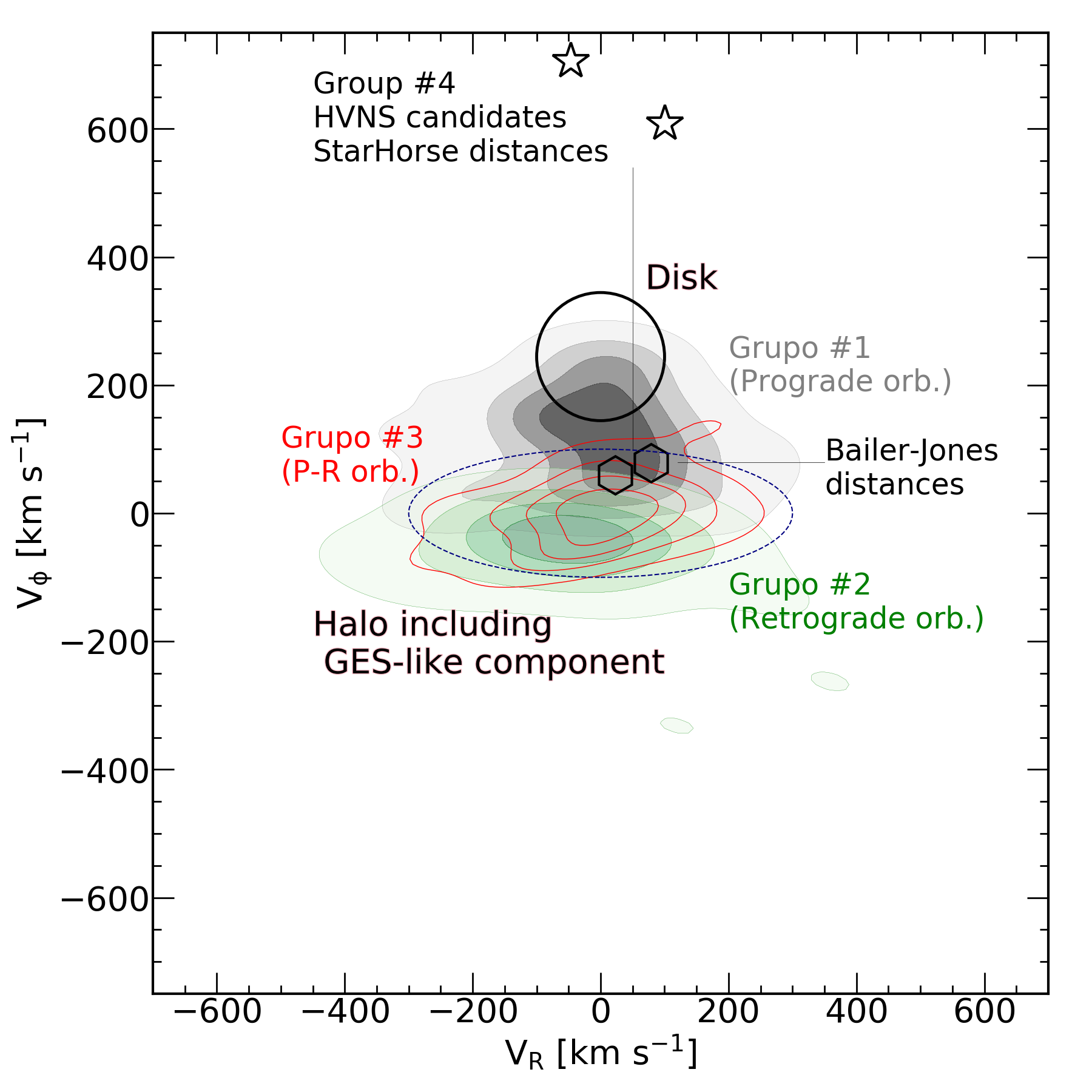}
                \caption{Same as Figure \ref{Figure2}(a), but here showing the KDE models of the newly identified N-rich stars confined to four main groups (see text). HVNS candidates (Grupo \#4; 2 stars) are shown as empty black star symbols for the assumed Bayesian \texttt{StarHorse} distances and as empty black hexagonal symbols by adopting the \texttt{Bailer-Jones} distances, while N-rich stars in prograde (Grupo \#1; 72 stars), retrograde (Grupo \#2; 25 stars), and P--R (Grupo \#3; 25 stars) orbital configurations are represented by the gray, green, and red contours, respectively. 
                }
                \label{Figure3}
        \end{center}
\end{figure}

In addition, the N-rich stars occupying the GES-like component in Figure \ref{Figure2} are dominated by a wide range of orbital configurations (including P--R, prograde and retrograde orbits), and they are almost completely confined to eccentricities larger than 0.65 peaking at $\sim$ 0.95, similar to that of the GES-like kinematic substructure identified by \citet{Naidu2020}. 

A fourth group (Group \#4) is made up of two N-rich outliers in the V$_{\rm \phi}$ versus V$_{\rm R}$ plane at $V_{\rm \phi}>600$ km s$^{-1}$. We refer to these stars as HVNS, and they likely form as part of the intriguing subgroup of the high-velocity stars \citep[][and references therein]{Hawkins2015}; readers are encouraged to refer to the discussion in Section \ref{HVNS}. 

Figure \ref{Figure2} also reveals that most of the metal-poor N-rich stars lie on high-eccentricity orbits with large vertical excursions above the Galactic plane, with some noticeable evidence of N-rich stars that likely formed in the outer halo, while the most metal-rich N-rich stars lie on prograde, low-eccentricity, inner halo- and disk-like orbital configurations. Interestingly, two (2M06243113$+$4213097 and 2M08534023$-$2554252) of the most metal-rich N-rich stars exhibit near circular orbital configurations in the outer disk of the MW, suggesting that they might have been formed in situ from material chemically self-enriched by a first generation of stars in the outer disk, or  they could have formed from a GC disruption in the inner Galaxy and radially migrated to their current location and/or are trapped by disk ripples \citep{Xu2015}. 

\section{Hypervelocity N-rich star (HVNS) candidates}
\label{HVNS}

Strikingly, we also identified two N-rich stars with a high azimuthal velocity of $V_{\rm \phi}>600$ km s$^{-1}$. Thus, these could be classified as atypical fast-moving N-rich stars likely escaping the MW potential, with the exception of the N-rich star 2M17025992$-$3537464 which exhibits a larger uncertainty in $V_{\rm \phi}\sim540$ km s$^{-1}$ overlapping with the kinematic of typical disk stars. We have classified these two stars as HVNS candidates. 
        
\begin{figure*}
        \begin{center}
                \includegraphics[width=180mm]{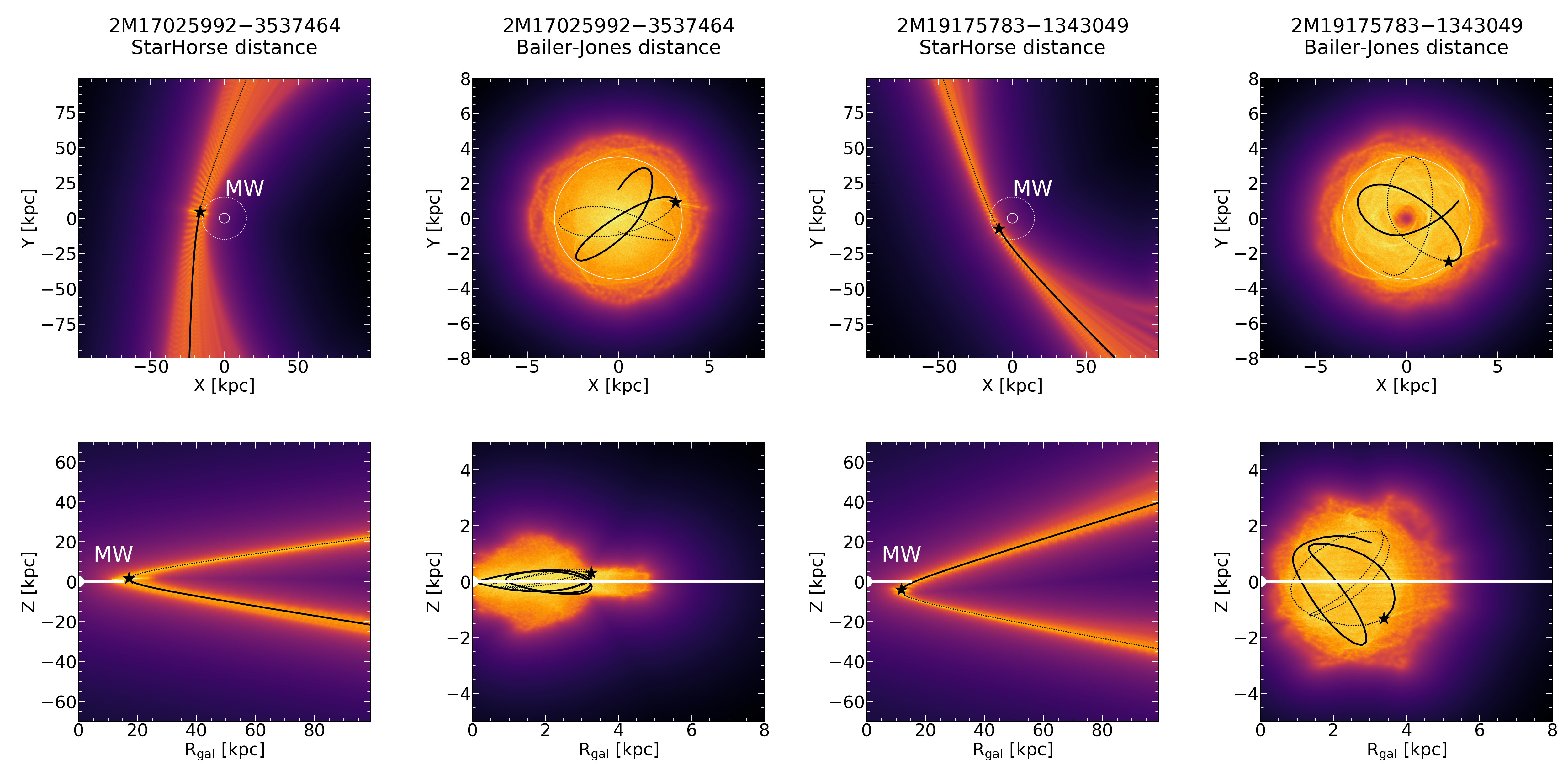}
                \caption{Ensemble of ten thousand orbits for the two HVNS candidates 2M17025992$-$3537464 and  2M19175783$-$1343049, projected on the equatorial (\textit{top}) and meridional (\textit{bottom}) Galactic planes in the inertial reference frame with a bar pattern speed of 41$\pm$10 km s$^{-1}$ kpc$^{-1}$,  and time-integrated forward and backward over 1 Gyr. The yellow and orange colors correspond to more probable regions of the space, which are  most frequently crossed by the simulated orbits. The black dashed and solid black lines show the 50$^{\rm th}$ percentile for the forward and backward orbital path, respectively. The top panels show, for guidance, a MW bulge (solid white circle) and disk (white dashed circle) radius of $\sim3.5$ kpc \citep{Barbuy2018} and $\sim$15 kpc, respectively. While  the MW disk extension until $\sim$15 kpc is also denoted by a thick white line in the bottom panels. Columns 1 and 3 show the orbital configurations by assuming the Bayesian \texttt{StarHorse} distances, whilst columns 2 and 4 display the predicted orbits by adopting the \texttt{Bailer-Jones} geometric distances. 
                }
                \label{Figure4}
        \end{center}
\end{figure*}

Figure \ref{Figure4} reveals that the predicted orbital configurations of these HVNS candidates are racing through space at above the escape speed of the MW, having a perigalactic distance of $r_{\rm peri}\sim11.7\pm0.87$ kpc (2M19175783$-$1343049) and $r_{\rm peri}\sim17.09\pm2.98$ kpc  (2M17025992$-$3537464), clearly indicating that they did not originate in the Galactic Center. However, their atypical chemistry suggests that these HVNS candidates likely have a GC formation site. 

These HVNS stars could have been produced via different physical processes, including the three-body encounters at the center of a GC; which through a process known as resonance scattering a binary star in the stellar cluster can capture a star in the cluster (thus forming a bound triple system), and after some time one of the stars (likely the HVNS) is ejected, or via flyby where a captured HVNS is ejected in the interaction \citep[see e.g.,][]{Hut1983, Pichardo2012, Fernandez-Trincado2016_47Tuc}. This finding could aid in providing useful information about the environments from which the high-velocity halo stars are produced. 
        
It is also important to note that \citet[hereafter \texttt{Bailer-Jones};][]{Bailer-Jones2021} provides estimated heliocentric distances for these two stars which put them closer to the Sun (d$_{\odot}< 6.5$ kpc) than those estimated with the Bayesian \texttt{StarHorse} code (d$_{\odot}>$19 kpc). Figures \ref{Figure2}--\ref{Figure4} reveal that adopting closer heliocentric distances for 2M19175783$-$1343049 and 2M17025992$-$3537464 puts these stars in an orbital configuration which inhabits the inner regions of the Galactic bulge in prograde and high eccentric ($\gtrsim$0.6) orbits, overlapping with the kinematic properties of the GES-like component ($V_{\rm \phi}<100$ km s$^{-1}$). 
        
However, the estimated distances from \citet{Bailer-Jones2021} for these two stars should be more uncertain as they should be dominated by a weak prior for a density distribution of the Galaxy, thus the priors would make us expect for these stars to be much closer than they truly are, producing unrealistic distances. Thus, it is likely that the distances from the \texttt{Bailer-Jones} catalog for 2M19175783$-$1343049 and 2M17025992$-$3537464 should just be following the priors. In addition, the Bayesian \texttt{StarHorse} code provides smaller uncertainties at larger ranges compared to the \texttt{Bailer-Jones} catalog. Therefore, we decided to use the Bayesian \texttt{StarHorse} distances derived by \citet{Queiroz2021} as the primary distance set for this work, also based on their better extinction treatment. However, with the upcoming \textit{Gaia} DR3, we will be able to constrain the distances and proper motions better for these two HVNS candidates.

\section{Conclusions}
 
We report on the discovery of 149 newly identified carbon-depleted ([C/Fe] $<+0.15$) N-rich stars over a wide range of metallicities ($-1.8<$ [Fe/H] $<-0.7$) for a grand total of $\sim$412 N-rich stars across the MW. Similar to the previously identified nitrogen-enhanced population, the new one exhibits elemental abundances comparable to those exclusively seen in GC environments. 

We also examined the position of the newly identified N-rich stars in the [Al/Fe] versus [Mg/Mn] plane as envisioned by \citet{Das2020}, finding that N-rich stars occupy a wide range at and beyond the locus deemed to contain an ``accreted" population ([Al/Fe] $\lesssim 0$ and [Mg/Mn] $\gtrsim+0.25$) and to that dominated by an in situ population of the MW. This suggests that the region occupied by the {in situ} population of the MW has a non-negligible presence of stars with unusual elemental abundances, likely associated to tidal stripping and/or the disruption of GCs over different epochs of evolution of the MW. The diverse dynamical and kinematical characteristics of the newly identified N-rich stars reveal that they do not share the same birthplaces in the MW, and that some of them could be the debris of GCs that were fully or partially destroyed during previous major merger accretion events such as the GES. 

        \begin{acknowledgements}  
                The author is grateful for the enlightening feedback from the anonymous referee.\\
                J.G.F-T gratefully acknowledges the grant support provided by Proyecto Fondecyt Iniciaci\'on No. 11220340, and also from ANID Concurso de Fomento a la Vinculaci\'on Internacional para Instituciones de Investigaci\'on Regionales (Modalidad corta duraci\'on) Proyecto No. FOVI210020.\\
                J.G.F-T and D.G.D. gratefully acknowledges the grant support from the Joint Committee ESO-Government of Chile 2021 (ORP 023/2021). \\
                T.C.B. acknowledges partial support for this work from grant PHY 14-30152: Physics Frontier Center / JINA Center for the Evolution of the Elements (JINA-CEE), awarded by the US National Science Foundation.\\
                B.B. acknowledges grants from FAPESP, CNPq and CAPES - Financial code 001. \\
                D.M. gratefully acknowledges support by the ANID BASAL projects ACE210002 and FB210003, and Fondecyt Project No. 1220724.\\
                E.R.G acknowledges support from ANID PhD scholarship No. 21210330.\\
                S.V. gratefully acknowledges the support provided by Fondecyt regular n. 1220264, and by the ANID BASAL projects ACE210002 and FB210003.\\
                D.G. gratefully acknowledges support from the ANID BASAL project ACE210002.\\
                D.G. also acknowledges financial support from the Direcci\'on de Investigaci\'on y Desarrollo de la Universidad de La Serena through the Programa de Incentivo a la Investigaci\'on de Acad\'emicos (PIA-DIDULS).\\
                 This work has made use of data from the European Space Agency (ESA) mission \textit{Gaia} (\url{http://www.cosmos.esa.int/gaia}), processed by the \textit{Gaia} Data Processing and Analysis Consortium (DPAC, \url{http://www.cosmos.esa.int/web/gaia/dpac/consortium}). Funding for the DPAC has been provided by national institutions, in particular the institutions participating in the \textit{Gaia} Multilateral Agreement.\\
                Funding for the Sloan Digital Sky Survey IV has been provided by the Alfred P. Sloan Foundation, the U.S. Department of Energy Office of Science, and the Participating Institutions. SDSS- IV acknowledges support and resources from the Center for High-Performance Computing at the University of Utah. The SDSS web site is www.sdss.org. SDSS-IV is managed by the Astrophysical Research Consortium for the Participating Institutions of the SDSS Collaboration including the Brazilian Participation Group, the Carnegie Institution for Science, Carnegie Mellon University, the Chilean Participation Group, the French Participation Group, Harvard-Smithsonian Center for Astrophysics, Instituto de Astrof\`{i}sica de Canarias, The Johns Hopkins University, Kavli Institute for the Physics and Mathematics of the Universe (IPMU) / University of Tokyo, Lawrence Berkeley National Laboratory, Leibniz Institut f\"{u}r Astrophysik Potsdam (AIP), Max-Planck-Institut f\"{u}r Astronomie (MPIA Heidelberg), Max-Planck-Institut f\"{u}r Astrophysik (MPA Garching), Max-Planck-Institut f\"{u}r Extraterrestrische Physik (MPE), National Astronomical Observatory of China, New Mexico State University, New York University, University of Notre Dame, Observat\'{o}rio Nacional / MCTI, The Ohio State University, Pennsylvania State University, Shanghai Astronomical Observatory, United Kingdom Participation Group, Universidad Nacional Aut\'{o}noma de M\'{e}xico, University of Arizona, University of Colorado Boulder, University of Oxford, University of Portsmouth, University of Utah, University of Virginia, University of Washington, University of Wisconsin, Vanderbilt University, and Yale University.\\
                This work has made use of data from the European Space Agency (ESA) mission \textit{Gaia} (\url{http://www.cosmos.esa.int/gaia}), processed by the \textit{Gaia} Data Processing and Analysis Consortium (DPAC, \url{http://www.cosmos.esa.int/web/gaia/dpac/consortium}). Funding for the DPAC has been provided by national institutions, in particular the institutions participating in the \textit{Gaia} Multilateral Agreement.\\   
    \end{acknowledgements}
        

\end{document}